\documentclass[prb,twocolumn,showpacs,preprintnumbers,amsmath,amssymb]{revtex4}
\usepackage{graphicx}% Include figure files
\usepackage{dcolumn}% Align table columns on decimal point
\usepackage{bm}% bold math
\begin{document}
%\preprint{APS/123-QED}
\title{Crossover of the dimensionality of 3$d$ spin fluctuations in LaCoPO}
\author{M. Majumder}
\author{K. Ghoshray}
\email{kajal.ghoshray@saha.ac.in}
\author{A. Ghoshray}
\author{B. Bandyopadhyay}
\author{B. Pahari}
\author{S. Banerjee}
\affiliation{Saha Institute of Nuclear Physics, 1/AF Bidhannagar, Kolkata-700064, India}
\date{\today}
\begin{abstract}
 dc magnetization and $^{31}$P spin lattice relaxation rate in the polycrystalline sample of LaCoPO suggest a spin fluctuation dominated ferromagnetically ordered state. Moreover, NMR data clearly indicate a crossover from 2D to 3D spin fluctuations across $T_{\mathrm{C}}$. In contrast to isotropic hyperfine field, $H_{\mathrm{hf}}$ at the $^{31}$P site in LaFePO, $H_{\mathrm{hf}}$ is anisotropic in LaCoPO. The data of spin lattice relaxation rate also exhibit anisotropic spin fluctuation. The anisotropy vanishes near $T_{\mathrm{C}}$.
\end{abstract}
\pacs{74.70.-b, 76.60.-k}
\maketitle
\section{Introduction}
Newly discovered correlated electron systems LnT$_{\mathrm{M}}$PnO [Ln=4$f$ rare earth element, T$_{\mathrm{M}}$=transition metal element with more than half-filled 3$d$ shell, Pn=pnictogen element] show interesting electronic and magnetic properties such as high transition temperature superconductivity, itinerant ferromagnetism, giant magnetoresistance, spin density wave (SDW) and structural instability \cite{Ishida09}. In particular, LnFeAsO, exhibit spin density wave (SDW) and structural instability \cite{Zhao,Martinelli09,McGuire08}. Moreover, they remain metallic to low temperature and only shows superconductivity ($T_{\mathrm{c}}$ in the range 26 - 55 K), when the SDW is suppressed towards zero temperature either through doping \cite{Kamihara08} or pressure \cite{Takahashi08}. In contrast, analogous phosphorous-based LaFePO, and LaNiPO are non magnetic metals and exhibit superconductivity ($T_{\mathrm{c}}$ in the range 2 - 6 K) due to strongly correlated electrons in the undoped form at ambient pressure, with the magnetic  ordering being suppressed due to the reduction of magnetic moments \cite{Hirano08,McQueen08,James08,Watanabe07,Tegel07}. It is be noted that $^{31}$P NMR results in (La$_{0.87}$Ca$_{0.13}$)FePO \cite{Nakai08} suggests the presence of short range ferromagnetic correlations with no long range magnetic ordering down to 2 K. However, in case of LaCoPO, the magnetic moment does not vanish completely due to odd number of electrons in 3$d$ orbitals. It has been shown that its ground state is itinerant ferromagnet with $T_C \approx$ 43 K, when measured in an external field of 0.1 T, with no superconducting transition down to 2 K \cite{Yanagi08}.

Recently, we presented $^{75}$As NMR study in partially oriented parent and F-doped CeFeAsO system focussing the importance of 4$f$ electron induced correlation effect over the obvious presence of the same due to 3$d$ electrons \cite{Ghoshray09}. In the present paper, we report the results of dc magnetization and $^{31}$P NMR in LaCoPO. $^{31}$P being a spin 1/2 nucleus, the resonance line shape would be affected only by the magnetic interaction and hence offer the opportunity to obtain unambiguous information about the low temperature electronic state. We have also performed $^{31}$P NMR study in LaFePO for comparison. The present results in LaCoPO suggest a strong field dependence on the onset (temperature) of ferromagnetic ordering suggesting the dominance of spin fluctuations even in the magnetically ordered state. Nuclear spin lattice relaxation ($T_1$) data also confirms this with an indication of a crossover of the dimensionality of the ferromagnetic spin fluctuations near 130 K.
\section{Results and discussion}
Polycrystalline samples of LaCoPO and LaFePO were synthesized by solid state reaction \cite{Yanagi08,Hirano08} and were characterized using powder x-ray diffraction at room temperature.
The magnetic moment was measured with a SQUID magnetometer (MPMSXL 7 T, Quantum Design). The NMR measurements were carried out at 7.04 T with a conventional phase-coherent spectrometer (Thamway PROT 4103). The powder sample of LaCoPO was magnetically aligned using epoxy
(Epotek-301).
%%%%%%%%%%%%%%%%%%%%%%%%%%%%%%%%%%%%%%%%%%%%%%%%%%%%%%%%%%%%%%%%%%%%
\begin{figure}[h]
{\centering {\includegraphics{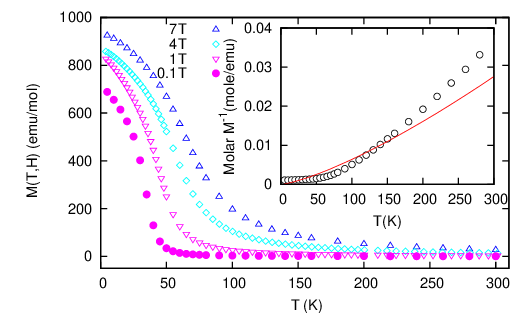}}\par}
\caption{(Color online) $M$ vs $T$ curve at 7 and 0.1 T. Inset: 1/$M$ vs $T$ curve; the solid line corresponds $T^{4/3}$.}
\label{susceptibility}
\end{figure}
%%%%%%%%%%%%%%%%%%%%%%%%%%%%%%%%%%%%%%%%%%%%%%%%%%%%%%%%%%%%%%%%%%%%

 The magnetic moment, $M$ vs $T$ curve (Fig. 1) for $H$=0.1 T shows sharp  enhancement below 50 K, due to the ferromagnetic ordering, as reported earlier \cite{Yanagi08}. As the NMR measurements were done in a field of 7 T, the bulk magnetization was measured in this field and at other intermediate fields in order to compare the bulk and the local magnetic properties derived from the NMR results. It is observed that, the enhancement of $M$ due to the magnetic ordering starts at a higher temperature depending on the field strength compared to that observed in a field of 0.1 T. Such a strong field dependence of T$_{\mathrm{C}}$ is also reported recently in layered compound La$_{1.2}$Sr$_{1.8}$Mn$_2$O$_7$ \cite{Hoch09}. One possible reason for this could be the existence of spin fluctuations in the ordered state. Application of a field of 7 T would reduce the effect of fluctuation and favor the ordering of the spins at much higher temperature. This is also supported by the observation of negative magnetoresistance in LaCoPO \cite{Yanagi08}.

%%%%%%%%%%%%%%%%%%%%%%%%%%%%%%%%%%%%%%%%%%%%%%%%%%%%%%%%%%%%%%%%
\begin{figure}[h]
{\includegraphics{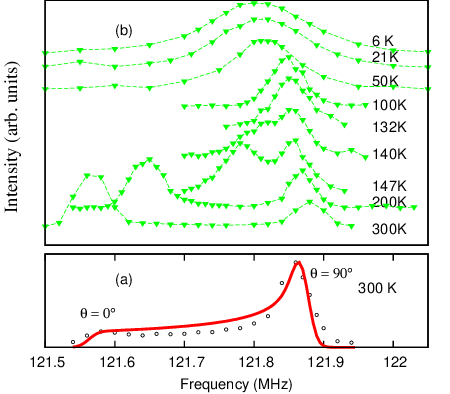}}
\caption{(Color online) (a): Powder pattern of LaCoPO (open circle) along with the calculated spectrum (continuous line). (b): $^{31}$P NMR spectra in partially aligned LaCoPO sample taken at 7.04 tesla, dotted line is for guide to eye.}
\label{spectra}
\end{figure}
%%%%%%%%%%%%%%%%%%%%%%%%%%%%%%%%%%%%%%%%%%%%%%%%%%%%%%%%%%%%
Fig. 2(a) shows a typical $^{31}$P NMR spectrum in polycrystalline LaCoPO which is anisotropic in nature. This corresponds to the powder pattern for a spin 1/2 nucleus with axially symmetric local magnetic field, as expected for tetragonal symmetry. The step in the low frequency side corresponds to $H\parallel c$ ($\theta=0^\circ$) and the peak in high frequency corresponds to $H\perp c$ ($\theta=\pi/2$).
The shift of the step with respect to the reference position ($\nu_R$), corresponds to $K_\parallel$
and that of the maximum corresponds to $K_{\perp}$. In Fig. 2(b) we have presented spectra of
partially oriented sample, where $\sim$60$\%$of the grains are oriented in the direction of $H\parallel$c.
As a result we have obtained two well resolved peaks in a same spectrum. Taking advantage of the partial alignment, $K_\parallel$ and $K_\perp$ can be determined accurately. This allows us to measure $K_{\mathrm{iso}}$ and $K_{\mathrm{ax}}$ from the relations $K_{\mathrm{iso}}$ = 2$K_{\mathrm{\perp}}$/3 + $K_{\parallel}$/3 and $K_{\mathrm{ax}}$ = 1/3($K_{\parallel}$ - $K_{\perp}$).
As the temperature is lowered the peak at $H\parallel$c shifts gradually towards the high frequency side. Whereas, the peak at $H\perp$c is shifted little towards the low frequency. Near 130 K, two peaks merge and start to broaden along with a small shift towards lower frequency. This temperature corresponds to the onset of the ferromagnetic transition in LaCoPO in a field of 7 T (Fig.1). Merging of the two peaks suggests the vanishing of the anisotropy of the local magnetic field $H_{local}$ near 130 K. A weak temperature dependence of the line position below 130 K along with a line broadening, is  expected for a magnetically ordered state. However, this broadening is not as high as it is in RFeAsO systems, where there is a huge enhancement of $^{75}$As NMR line width in the ordered state due to the spin density wave (SDW) transition. Whereas, in LaCoPO, the spectrum can be detected even at 4 K, well below T$_{\mathrm{C}}$, suggesting a homogeneous ordered magnetic field along with an itinerant character of the Co 3$d$ electrons responsible for the ordering. Fig. 3(a) shows the variation of $K_{\parallel}$ and $K_{\perp}$ as a function of temperature.

%%%%%%%%%%%%%%%%%%%%%%%%%%%%%%%%%%%%%%%%%%%%%%%%%%%%%%%%%%%%%%%%%%%%%%%%%%%%%%
\begin{figure}[h]
\includegraphics{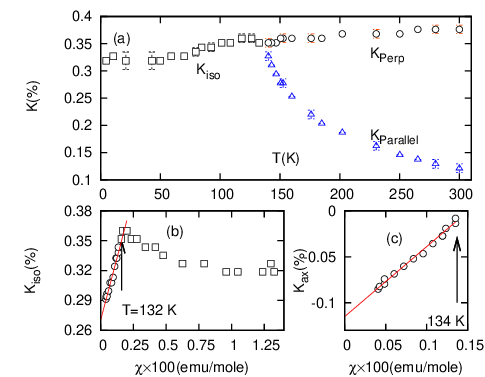}
\caption{(Color online) (a): Temperature dependence of $^{31}$P NMR shift. (b), (c): Represent variation of
$K_{\mathrm{iso}}$ and $K_{\mathrm{ax}}$ with $\chi$ respectively. The lines correspond linear fit.}
\label{shift}
\end{figure}
%%%%%%%%%%%%%%%%%%%%%%%%%%%%%%%%%%%%%%%%%%%%%%%%%%%%%%%%%%%%%%

The shift is mainly governed by the two contributions: (1) temperature independent $K_0$ due to conduction electrons, and (2) temperature dependent $K(T)$ due to  localized character of the d-electrons. So, $K_{\mathrm{total}}$ = $K_0$ + $K(T)$, where
 \begin {equation}
 K(T) =H_{\mathrm{hf}}^d\chi(T)/N_A\mu_B,
 \end {equation}
where $H_{\mathrm{hf}}^{\mathrm{d}}$ is the hyperfine field per  $\mu_B$. As long as $H_{\mathrm{hf}}^{\mathrm{d}}$ remains unchanged, $K(T)$ is proportional to $\chi(T)$. Fig. 3(b) and (c) show the variation of $K_{\mathrm{iso}}$ and $K_{\mathrm{ax}}$ with $\chi$. Both the parameters show a linear variation in the range 130 - 300 K, with  $H_{\mathrm{hf}}^{\mathrm{iso}}$ = 2.79 kOe/$\mu_B$ and $H_{\mathrm{hf}}^{\mathrm{ax}}$ = -4.24 kOe/$\mu_B$ The value of $H_{\mathrm{hf}}^{\mathrm{iso}}$ is found to be much smaller than that reported in case of Ca- doped LaFePO ($H_{\mathrm{hf}}$=12.5 kOe/$\mu_B$) \cite{Nakai08}. A significant deviation of the $K_{\mathrm{iso}}$ vs $\chi$ plot linearity, below 130 K, indicates a modification of the electronic wave function contributing to the
hyperfine coupling, responsible for the temperature dependent shift. There are two possibilities; one is the partial delocalization of the $d$-electrons due to mixing with the $s$-band near this temperature range, and the other is the crystal field effect.
\begin{figure}
\includegraphics{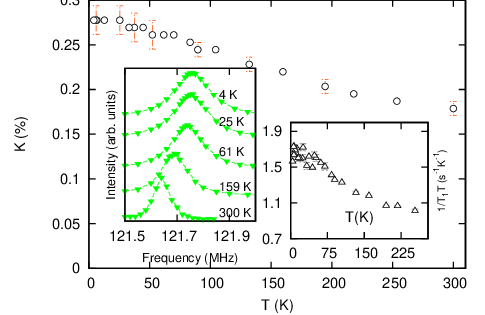}
\caption{(Color online) Variation of $^{31}$P shift with $T$ in LaFePO. Inset: $^{31}$P NMR spectra and T vs $1/T_1T$ in LaFePO.}
\label{shiftFe}
\end{figure}
%%%%%%%%%%%%%%%%%%%%%%%%%%%%%%%%%%%%%%%%%%%%%%%%%%%%%%%%%%

To compare the effect of complete replacement of Co by Fe, $^{31}$P NMR study in isostructural LaFePO was also performed. In this case a symmetric resonance line (Inset of Fig. 4) was observed in the range 4-300 K with slow and gradual increase in line-width and shift (Fig. 4), and both of these parameters become almost $T$ independent below 50 K, as was reported earlier in Ca doped LaFePO \cite{Nakai08}.
This $T$ independent behavior of $K$ below 50 K in LaFePO (which does not show ordering), is very similar to that in LaCoPO in the ordered state.

To verify the suggestion \cite{Yanagi08} that LaCoPO is an itinerant ferromagnet and the magnetic property should be governed by the spin fluctuations,
we have measured $T_1$ at the peaks
corresponding to H$\parallel$c and H$\perp$c. The recovery of nuclear magnetization in each case follows a single exponential as expected for a spin-1/2 nucleus. This confirms the absence of any impurity contribution in $^{31}$P NMR. Inset of Fig. 5 shows the recovery curve for the peak at $H\parallel$c as a function of the product of the delay time ($\tau)$ and T. It clearly shows that $T_1T$ decreases in the range 150 - 300 K and then increases till 100 K and stays constant down to 4 K, signifying a metallic character.

%%%%%%%%%%%%%%%%%%%%%%%%%%%%%%%%%%%%%%%%%%%%%%%%%%%%%%%%%%%%%%%
\begin{figure}[h]
\includegraphics{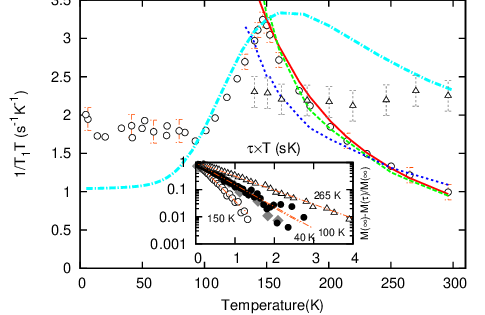}
\caption{(Color online) T dependence of $1/T_1T$ in LaCoPO; $\bigcirc: (1/T_1T)_{H\|c}$, and $\bigtriangleup: (1/T_1T)_{H\perp c}$. The dashed and dotted curves correspond to $\chi^{3/2}$ and $\chi^{1}$ respectively as defined in Eq. 3.  continuous curve correspond to Eq. 4. The dot-dashed curve  represents Eq. 5. Inset: Recovery curves vs $\tau$ $\times$ T at different temperatures in LaCoPO, fitted with a single exponential.}
\label{rate1}
\end{figure}
%%%%%%%%%%%%%%%%%%%%%%%%%%%%%%%%%%%%%%%%%%%%%%%%%%%%%%%%%%

$(1/T_1T)_{H\|c}$ (Fig. 5) shows a faster increment in the range 150 - 300 K compared to that in LaFePO, indicating the effect of slowing down of the 3$d$ spin fluctuations due to the development of short range correlations above the magnetic transition temperature. The peak near 150 K suggests the development of long range correlations below this temperature. This gradually reduces the contribution of 3$d$ spin fluctuations to the $^{31}$P nuclear relaxation process, below 150 K. Finally an almost $T$ independent behavior below 100 K signifies the dominant role of the Korringa contribution due to $s$-electrons in the relaxation process and a less significant role of 3$d$ spin fluctuations, as expected in a magnetically ordered state. Thus the present NMR result confirm microscopically the occurrence of a long range magnetic order in LaCoPO at a higher temperature in a field of 7 T, compared to that reported from magnetic susceptibility data at H=0.1 T. For the peak at H $\perp$ c, $(1/T_1T)_{H\perp c}$, shows a temperature independent Korringa behavior, similar to that of $K_\perp$. Thus in LaCoPO, both the Knight shift and $1/T_1$ are anisotropic in the range 132-300 K.

$(1/T_1)_\alpha$ ($\alpha$=a,b,c) is proportional to the fluctuations of the local magnetic field  $\delta h^2$ perpendicular to the $\alpha$ axis. Thus, $(1/T_1)_{H\parallel c}\propto[\delta h_a^2+ \delta h_b^2]$, $(1/T_1)_{H\perp c}\propto[\delta  h_a^2+\delta  h_c^2\rangle]$. The P site being axially symmetric, $\delta h_a^2=\delta h_b^2$. Since the contribution to $(1/T_1T)_{H\parallel c}$ comes from the fluctuations of the hyperfine field $\perp$ to $c$-axis, the $T$ dependence of $(1/T_1T)_{H\parallel c}$ indicates a slowing down of the Co-3$d$ spin fluctuations in the $a$-$b$ plane. On the other hand,  $(1/T_1T)_{H\perp c}$ probes the fluctuations of the local field in $c$-direction. So its $T$ independent behavior suggests that the Co-3d spin fluctuations in LaCoPO, is confined within the $a$-$b$ plane. Thus $(1/T_1T)_{H\perp c}$ is governed only by the scattering due to conduction electrons.

\subsection{Nature of spin fluctuation}
In order to understand the nature of the magnetic correlations we have analyzed the $(1/T_1)_{H\parallel c}$ data using the theoretical models for the electron spin fluctuation (SF) contribution.
In general (1/$T_1T)_{\mathrm{SF}}$ is given by \cite{Nowak09}
 \begin {equation}
(1/T_1T)_{\mathrm{SF}} \propto (\gamma_n A_{\mathrm{hf}})^2\sum_q\chi\prime\prime(q,\omega_n)/\omega_n,
 \end {equation}
where $\chi\prime\prime(q,\omega_n$) is the imaginary part of the transverse dynamical electron-spin susceptibility, $\gamma_n$ and $\omega_n$ are the nuclear gyromagnetic ratio and Larmor frequency, respectively.
When 2D(3D) spin fluctuations are dominant \cite{Ishigaki98,Hatatani95,Nakai08},
\begin {equation}
  1/T_1T \propto\chi(q=0)^{3/2(1)}.
 \end {equation}

 In the present case, $K$ is proportional to $\chi$ in the range 150 - 300 K (Fig. 3a), so in calculation of $1/T_1T$ using Eq. 3, we have used $K$, since the Knight shift probes the intrinsic spin susceptibility of the sample. It is seen from Fig. 5 that the experimental data agree satisfactorily with the calculated (continuous) curve of 2D spin fluctuation, but theoretical (dotted) curve of 3D spin fluctuation can not be fitted to the experimental data, suggesting the predominance of ferromagnetic 2D spin fluctuations in the paramagnetic phase. According to the extension of self consistent renormalization (SCR) theory for 2D itinerant ferromagnetic metal near magnetic instability \cite{Hatatani95}
  \begin {equation}
  1/T_1T \propto T^{-3/2}(-lnT)^{-3/2}.
  \end {equation}
The calculated (dashed) curve also matches well with the experimental results (Fig. 5). This finding also corroborates with the above conclusion.

On the other hand, according to  (SCR) theory of spin fluctuations, for weak itinerant ferromagnet (WIF), the ferromagnetic (q = 0) 3D spin-fluctuation contribution to $1/T_1T$ in presence of magnetic field, both in the paramagnetic and the ferromagnetic region is given by \cite{Moriya,Rabis05,Yoshimura87}
 \begin {equation}
 1/(T_1T)_{\mathrm{SF}} = k\chi/(1+\chi^3H^2P) + \beta,
 \end {equation}
where $P$ is a constant related to the area of the fermi surface of the magnetic electrons and $k$ is related to the energy width of the dynamical spin-fluctuation spectrum. $\beta$ is temperature independent and contains the sum of the contributions due to orbital moments of $p$ and $d$ electrons, Fermi contact contribution of $s$ conduction electrons and that due to the spin dipolar interaction with $p$ and $d$ electrons. Fig. 5 shows that the experimental curve deviates from Eq. 5 (dot-dashed curve) above 150 K and agrees in the range 80 - 150 K. So this relation is not satisfied in both the paramagnetic and the ordered state as it should \cite{Yoshimura87}. It is seen that even in the ordered state, below 80 K the calculated curve lies below the experimental curve, while the nature of both the curves remain same. A possible reason for this could be the presence of a small amount of non magnetic impurity, which would reduce the magnitude of the measured bulk magnetic moment used for calculating 1/T$_1$T using Eq. 5. Since the nuclear relaxation time probes the intrinsic magnetic property of a system, the 1/T$_1$T data is expected to be more accurate compared to those calculated using bulk magnetization data, particularly at low temperatures. Furthermore, the anisotropic nature of the relaxation time reduces to zero i.e. $(1/T_1T)_{H\|c} = (1/T_1T)_{H\perp c}$, near the temperature where the Knight shift anisotropy also vanishes. This suggests that below $T_C$ the spin fluctuation is 3D in nature. Since the spin fluctuation is 2D in nature above 150 K, the magnitude of $T_1$ should be greater than that in the case where 3D spin fluctuation dominates. Possibly this is the reason for the smaller value of $T_1$ obtained for the calculated value of $1/T_1T$  using Eq. (5) above 150 K.

Moreover, the SCR theory of WIF with 3D spin fluctuatons\cite{Moriya,Ohta09}, predicts that 1/$\chi$ or 1/$M$ should vary as $T^{4/3}$. However it can be seen from the inset of Fig. 1 that above 150 K 1/$\chi$ deviates from this relation which indicates that there is a change in the nature of the spin fluctuation near $T_C$. Thus from magnetization and the relaxation data, we conclude that their is a crossover from 2D to 3D ferromagnetic spin fluctuations across 150 K. Below 100 K, the relaxation process is mainly governed by the Korringa process.
The $1/T_1T$ vs $T$ curve for LaFePO is shown in the inset of Fig. 4. The behavior is very similar to that reported in Ca-doped LaFePO.The slow but gradual increase of $1/T_1$ with $T$ in the range 50 - 300 K
could be a signature of the weaker short range correlations among the 3d spins of Fe in LaFePO compared to that in LaCoPO.

Contribution of the anisotropy of $K$ and $1/T_1T$ arises from both the dipolar and hyperfine fields. The former is proportional to the bulk susceptibility($\chi$). As in case of LaCoPO, $\chi$ increases in the range 135-300, so a decrease in the $^{31}$P line shape anisotropy in this range, clearly suggests a decrease in the anisotropy of the hyperfine field, with a negligible contribution from the dipolar field anisotropy.
\section{Conclusion}
The study of dc magnetization in LaCoPO in the range 4 - 300 K in presence of different magnetic fields, suggest the dominance of 3$d$ spin fluctuations in the ferromagnetically ordered state. This is further confirmed by the $^{31}$P nuclear $T_1$ measurements. The $T$ dependence of 1/$T_1T$ in LaCoPO suggests the dominance of 2D ferromagnetic spin fluctuations in the paramagnetic phase with a crossover to the 3D ferromagnetic spin fluctuation regime near the ordering temperature. Moreover, both the static and the dynamic part of the hyperfine fields are anisotropic in the paramagnetic phase and each of them becomes isotropic in the ordered phase, where the bulk susceptibility (at 7 T) shows a sharp enhancement, due to the ferromagnetic ordering. In contrast, $^{31}$P $H_{\mathrm{hf}}$ is isotropic in LaFePO in the range 4 - 300 K. Electronic structure calculation \cite{Hirano08,Yanagi08} for both compounds show similar Fermi surface composed of hybridized P 3$p_z$, Fe/Co 3$d_{z^2}$ and 3($d_{xz},d_{yz}$) orbitals. The only probable difference in LaCoPO because of the presence of odd 3d-electrons (3d$^7$) is the participation of the upper Fe 3$d_{x^2 - y^2}$ orbital along with the 3$d_{z^2}$ states in the Fermi level inducing the magnetic moments in LaCoPO. More insight of the spin fluctuation might be revealed by studying NMR with a single crystal at different magnetic field.

\end{document}